\documentclass[preprint,11pt]{elsarticle}   
\usepackage{amssymb}
   
\journal{Computer Physics Communications}

\def\code{{\tt SUSY\_FLAVOR}}

\begin{document}
\begin{frontmatter}

\title{SUSY FLAVOR v2.5: a computational tool for FCNC and
  CP-violating processes in the MSSM}

\author[1]{J. Rosiek\fnref{aut2}}\address[1]{Institute of Theoretical
  Physics, Physics Department, University of Warsaw\\[-40mm]
  \rightline{\rm{\large IFT-14-04}}\vskip 30mm }

\fntext[aut2]{janusz.rosiek@fuw.edu.pl}

\begin{abstract}
We present \code{} version 2.5 --- program that calculates over 30
low-energy flavor observables in the general $R$-parity conserving
MSSM.  Comparing to previous versions, in \code{} v2.5 parameter
initialization in SLHA2 formats has been significantly generalized, so
that the program accepts most of the output files produced by other
libraries analyzing the MSSM phenomenology. Number of bugs and
inconsistencies have been fixed, based on users feedback.
Calculations of several processes implemented in earlier versions have
been corrected. New processes of rare decays of the top quark to Higgs
boson have been included.  Variables controlling inclusion of
contributions from various MSSM sectors have been added. Full updated
manual of \code{} v2.5 integrating the details of the modifications
listed below can be found at {\tt arxiv.org/abs/arXiv:1203.5023}.
\end{abstract}

\end{frontmatter}

\noindent
{\bf NEW VERSION PROGRAM SUMMARY}\\[2mm] 
{\em Program Title:} SUSY FLAVOR
v2.5\\
{\em Authors:}~ J.~Rosiek, P.~Chankowski, A.~Crivellin, A.~Dedes,
S.~J\"ager, P.~Tanedo\\
%
%
%
{\em Programming language:} Fortran 77.\\
{\em Computer:} Any.\\
{\em Operating system:} Any, tested on Linux.
{\em Classification:} 11.6.\\
{\em Catalogue identifier of previous version:} AEGV\_v2\_0\\
{\em Journal reference of previous version:} Comput. Phys. Commun. 184
(2013) 1004\\
{\em Does the new version supersede the previous version?:} Yes\\
{\em Nature of problem:}\\ Predicting CP-violating observables, meson
mixing parameters and branching ratios for set of rare processes in
the general R-parity conserving MSSM.\\
{\em Solution method:}\\ We use standard quantum theoretical methods
to calculate Wilson coefficients in MSSM and at one loop including QCD
corrections at higher orders when this is necessary and possible. \\
{\em Reasons for the new version:}\\ The input/output routines have
been rewritten to make them more flexible and compatible with the
SLHA2 standard~\cite{SLHA2}. Calculations of the several processes
implemented in earlier \code{} versions have been corrected.  New
observables have been added. Number of bugs have been corrected.\\
{\em Summary of revisions:}\\
1. Modified initialization routines. Currently the program should be
able to read without modifications most of the SLHA2-compatible output
files produced by other publicly available libraries calculating
observables related to the MSSM phenomenology. In addition, new
optional input block {\tt SFLAV\_HADRON} has been defined to
facilitate modifications of the parameters related to the hadronic and
QCD sector.

The initialization sequence goes now through the following steps:
\begin{itemize}
\item Before reading the file, all parameters are set to some initial
  values (which can be changed by editing the values given in the
  subroutine {\tt sflav\_defaults} in file {\tt sflav\_io.f}).
\item Subsequently, the user-defined data are read from the file with
  the default name {\tt susy\_flavor.in}.  Data are grouped in Blocks
  following the SLHA2 specification or extensions described
  in~\cite{SFLAV2}. Blocks are read in the following order: {\tt
    SOFTINP}, {\tt SMINPUTS}, {\tt VCKMIN}, {\tt MINPAR} ($\tan\beta$
  only, other entries are ignored), {\tt EXTPAR}, {\tt IMEXTPAR}, {\tt
    MSL2IN}, {\tt IMMSL2IN}, {\tt MSE2IN}, {\tt IMMSE2IN}, {\tt TEIN},
  {\tt IMTEIN}, {\tt TEINH}, {\tt IMTEINH}, {\tt MSQ2IN}, {\tt
    IMMSQ2IN}, {\tt MSU2IN}, {\tt IMMSU2IN}, {\tt MSD2IN}, {\tt
    IMMSD2IN}, {\tt TUIN}, {\tt IMTUIN}, {\tt TUINH}, {\tt IMTUINH},
  {\tt TDIN}, {\tt IMTDIN}, {\tt TDINH}, {\tt IMTDINH}, {\tt
    SFLAV\_HADRON}.
\item Presence of {\em any} Block is optional - if some Block is
  absent, program falls back to default parameter values.  At least
  flavor-diagonal SUSY mass parameters have to be defined, otherwise
  the vanishing default values cause program to crash.
\item If a parameter is multiply defined in several Blocks, the value
  from Block read as latest in the list above overwrites (without
  warning!) the values from preceding Blocks.
\item Blocks do not need to be complete and to contain all entries
  described in the SLHA2 specification - it is sufficient to define a
  minimal set of the parameters relevant for a given problem, others
  are filled with the default values.
\item The ``non-holomorphic'' LR mixing terms are not included in the
  SLHA2 specification and by default set to 0.  They can be
  initialized to the non-trivial values in the blocks {\tt TXINH} and
  {\tt IMTXINH} ($X=E,D,U$)
\item New input block {\tt SFLAV\_HADRON} allows to modify hadronic-
  and QCD-related quantities used by \code.  The structure of this
  block and the default values of hadronic parameters are shown at the
  end of sample input file {\tt susy\_flavor.in} attached to the
  \code{} distribution.
\end{itemize}
2. New control variables has been added, allowing to separately switch
contributions from various MSSM sectors on or off.  They can be set by
the following statement at the beginning of the driver program:

 {\tt call set\_active\_sector(ih,ic,in,ig)},

\noindent where the variables {\tt ih, ic, in, ig} can take values $0$
or $1$ and they control, respectively, the inclusion in the total
result the diagrams with gauge and Higgs bosons, charginos,
neutralinos and gluinos exchanged in the loops. Note that diagrams
with Higgs and gauge bosons are always added together and currently
cannot be disentangled, so setting {\tt ih=1, ic=in=ig=0} does not
reproduce the SM result.  By default, if no call to {\tt
  set\_active\_sector} is made, all control variables are assumed to
be equal to 1, so that all contributions are included.\\[2mm]
3. Added or modified processes:
\begin{itemize}
\item The expressions used to calculate the neutron Electric Dipole
  Moment have been modified.
\item The branching ratios for the radiative decays of the heavy
  lepton into the lighter lepton and the photon, $ \mu\to e\gamma$ and
  $\tau\to e\gamma$, $\mu\gamma$, are now normalized to the total
  heavy lepton decay width (previously they were normalized to the
  decay width into leptonic channels).
\item The routines calculating branching ratios of $B\to\tau\nu$ and
  $B\to D\tau\nu$ decays have been generalized to include more general
  structure of the effective Higgs boson-fermion couplings. In
  addition the routine calculating $Br(B\to D^\star\tau\nu)$ has been
  added.
\item The routines for rare decays of the top quark to the CP-even
  Higgs boson and the light quarks, $t\to ch,uh$, have been added,
  based on Ref.~\cite{T2UH} (program can calculate also the decay
  rates of the top quark to the heavier CP-even Higgs boson $H$,
  assuming that such decays are allowed kinematically).
\item The routine calculating the approximate 2-loop estimate of the
  neutral CP-even Higgs mass $m_h$ has been added, based on
  Ref.~\cite{Heinemeyer:1999be}. Note that for the more precise
  calculations of this mass other publicly available SUSY codes should
  be used.
\item Default values of numerous quantities which are treated by
  \code{} as the external parameters, mainly the values of hadronic
  parameters obtained from lattice calculations and results of
  experimental measurements, have been updated to accommodate the
  latest published results.
\end{itemize}
4. \code's output is now written to the file named {\tt
  susy\_flavor.out}.  It has ``SLHA-like'' structure, i.e. it is
divided into ``data blocks'', however these blocks are \code{}
specific and do not follow the common SLHA2 standards.  The output
file contains the following data blocks:
\begin{itemize}
\item
{\tt SFLAV\_CONTROL}: control variables and error code status.
\item {\tt SFLAV\_MASS}: MSSM mass spectrum after mass matrix
  diagonalization.
\item {\tt SFLAV\_CHIRAL\_YUKAWA}: relative size of the resummed
  chiral corrections to the Yukawa couplings.
\item {\tt SFLAV\_CHIRAL\_CKM}: relative size of the resummed chiral
  corrections to the CKM matrix elements.
\item {\tt SFLAV\_DELTA\_F0}: $\Delta F = 0$ observables: leptonic
  EDMs and $g-2$ anomalies, neutron EDM.
\item {\tt SFLAV\_DELTA\_F1}: $\Delta F = 1$ observables: decay rates
  of $l\to l'\gamma$, $K\to\pi\bar\nu\nu$, $B^+ \to \tau^+ \nu$, $B
  \to D \tau \nu$, $B \to D^\star \tau \nu$, $B \to X_s \gamma$,
  $B_{d,s}\to l^+_i l_j^-$, $t\to uh$, $t\to ch$.
\item {\tt SFLAV\_DELTA\_F2}: $\Delta F = 2$ observables:
  $\epsilon_K$, $\Delta m_K$, $\Delta m_D$, $\Delta m_{B_d}$, $\Delta
  m_{B_s}$.
\end{itemize}
Blocks {\tt SFLAV\_CHIRAL\_YUKAWA} and {\tt SFLAV\_CHIRAL\_CKM} show
the relative differences of bare and physical Yukawa couplings and CKM
matrix elements after the resummation of chiral corrections. If they
are large, $\geq {\cal O}(1)$, the perturbation expansion is not
reliable and the remaining program output may not be correct.\\[2mm]
5. The full updated manual for \code{} v2.5 has been created,
integrating the detailed description of the modifications listed
above. It can be downloaded at the address {\tt
  arxiv.org/abs/arXiv:1203.5023}.  Regular code distribution updates
and bug fixes (between the major revisions submitted to Computer
Physics Communications) can be found on the program web page {\tt
  www.fuw.edu.pl/susy\_flavor}.\\[2mm]
{\em Restrictions:} The results apply only to the case of MSSM with
R-parity conservation and without heavy right neutrino
sector~\cite{DEHARO}.\\
{\em Running time:} For a single parameter set below 1s on a personal
computer.

\subsection*{Acknowledgements}

\noindent 
The author would like to thank to A. Dedes, M. Paraskevas and K. Suxho
for collaboration in the calculations of the rare top quark decay
rates.  This work was supported in part by the Polish National Science
Center under the research grants DEC-2011/01/M/ST2/02466 and
DEC-2012/05/B/ST2/02597.

\end{document}